\documentclass[reprint,%showpacs,
preprintnumbers,amsmath,amssymb]{revtex4}
\usepackage{graphicx}% Include figure files
\usepackage{dcolumn}% Align table columns on decimal point
\usepackage{bm}
\numberwithin{equation}{section}

\usepackage{hyperref}

\numberwithin{equation}{section}

\newcommand{\bee}{\begin{equation}}
\newcommand{\ene}{\end{equation}}

\newcommand{\pa}{\partial}

\begin{document}

\title{Traveling Wavetrains in the Complex Cubic--Quintic Ginzburg--Landau Equation}

\author{Stefan C. Mancas}
\email{mancass@erau.edu}
\affiliation{Department of Mathematics, Embry-Riddle Aeronautical University, Daytona-Beach, FL 32114-3900, USA}

\author{S. Roy Choudhury}
\email{choudhur@longwood.cs.ucf.edu}
\affiliation{Department of Mathematics, University of Central Florida, Orlando, FL 32816-1364}

\begin{abstract}

\vspace{.1in}

In this paper we use a traveling wave reduction or a so--called spatial approximation to comprehensively investigate the periodic solutions of the complex cubic--quintic Ginzburg--Landau equation. The primary tools used here are Hopf bifurcation theory and perturbation theory. Explicit results are obtained for the post--bifurcation periodic orbits and their stability. Generalized and degenerate Hopf bifurcations are also briefly considered to track the emergence of global structure such as homoclinic orbits.
\end{abstract}
\maketitle

\section{Introduction}
The cubic complex Ginzburg--Landau equation (CGLE) is the canonical equation governing the weakly nonlinear behavior of dissipative systems in a wide variety of disciplines \cite{Dodd}. In fluid mechanics, it is also often referred to as the Newell--Whitehead equation after the authors who derived it in the context of B\'enard convection \cite{Dodd,Drazin:1}.

As such, it is also one of the most widely studied nonlinear equations. Many basic properties of the equation and its solutions are reviewed in \cite{Aranson,Bowman}, together with applications to a vast variety of phenomena including nonlinear waves, second--order phase transitions, superconductivity, superfluidity, Bose--Einstein condensation, liquid crystals and string theory. The numerical studies by Brusch et al. \cite{Brusch:1,Brusch:2} which primarily consider periodic traveling wave solutions of the cubic CGLE, together with secondary pitchfork bifurcations and period doubling cascades into disordered turbulent regimes, also give comprehensive summaries of other work on this system. Early numerical studies \cite{Keefe,Landman} and theoretical investigations \cite{Newton:1,Newton:2} of periodic solutions and secondary bifurcations are also of general interest for our work here.

Certain situations or phenomena, such as where the cubic nonlinear term is close to zero, may require the inclusion of higher--order nonlinearities leading to the so--called cubic--quintic CGLE. This has proved to be a rich system with very diverse solution behaviors. In particular, a relatively early and influential review by van Saarloos and Hohenberg \cite{Saarloos}, also recently extended to two coupled cubic CGL equations \cite{Hecke,Alvarez}, considered phase--plane counting arguments for traveling wave coherent structures, some analytic and perturbative solutions, limited comparisons to numerics, and so--called ``linear marginal stability analysis'' to select the phase speed of the traveling waves.

Among the multitude of other papers, we shall only refer to two sets of studies which will directly pertain to the work in this thesis. The first class of papers \cite{Holmes,Doelman:1,Doelman:2} and \cite{Duan,Doelman:3} used dynamical systems techniques to prove that the cubic--quintic CGLE admits periodic and quasi--periodic traveling wave solutions.

The second class of papers \cite{Soto,Akhmediev:1}, primarily involving numerical simulations of the full cubic--quintic CGL PDE in the context of Nonlinear Optics, revealed various branches of plane wave solutions which are referred to as continuous wave (CW) solutions in the Optics literature. More importantly, these latter studies also found various spatially confined coherent structures of the PDE, with envelopes which exhibit complicated temporal dynamics. In \cite{Akhmediev:1}, these various structures are categorized as plain pulses (or regular stationary solutions), pulsating solitary waves, creeping solitons, slugs or snakes, erupting solitons, and chaotic solitons depending on the temporal behavior of the envelopes. In addition, note that the speed of the new classes of solutions may be zero, constant, or periodic (since it is an eigenvalue, the speed may be in principle also quasiperiodic or chaotic, although no such cases appear to have been reported). All indications are that these classes of solutions, all of which have amplitudes which vary in time, do not exist as stable structures in Hamiltonian systems. Even if excited initially, amplitude modulated solitary waves restructure into regular stationary solutions \cite{Artigas}. Exceptions to this rule are the integrable models where the pulsating structures are nonlinear superpositions or fundamental solutions \cite{Satsuma}. Hence, these classes of solutions are novel and they exist only in the presence of dissipation in the simulations of \cite{Akhmediev:1}. Also, secondary complete period doubling cascades of the pulsating solitons leading as usual to regimes of chaos are also found. This last feature for numerical solutions of the full cubic--quintic PDE is strongly reminiscent of the period doubling cascades found in \cite{Brusch:1,Brusch:2} for period solutions of the traveling wave reduced ODEs of the cubic CGLE.

Motivated by the above, we begin a fresh look at the traveling wave solutions of the cubic--quintic CGLE in this paper. Besides attempting to understand the complex numerical coherent structures in \cite{Akhmediev:1}, one other goal is to build a bridge between the dynamical systems approach in \cite{Holmes,Doelman:1,Doelman:2}--\cite{Duan,Doelman:3} and the numerical one in \cite{Soto,Akhmediev:1}. Given the importance of the cubic--quintic CGLE as a canonical pattern--forming system, this is clearly important in and of itself. However, a word of warning is in order here. Some of the features in \cite{Akhmediev:1} may well be inherently spatio--temporal in nature, so that a spatial traveling--wave reduction may not be sufficient to completely capture all aspects. Indeed, there is some evidence along these lines \cite{Artigas}.

In this paper we begin by using a traveling wave reduction or a so--called spatial approximation to investigate the periodic solutions of the CGLE. The primary tool used here is Hopf bifurcation theory. Immediately following the Hopf bifurcations we construct the periodic orbits by using the method of multiple scales.

The remainder of the paper is organized as follows. We first analyze the stability of fixed points in Section 2  and the onset of instability via a Hopf bifurcation, which may be either supercritical or subcritical. Then stability of periodic orbits is presented in Section 3 where we derive analytical expressions for the periodic orbits resulting from this Hopf bifurcation, and for their stability coefficients, by employing the multiple scales method. Section 4 considers numerical solutions and discusses the results. Generalized and degenerate Hopf bifurcations have also  been considered to track the emergence of global structure such as homoclinic orbits in Section 5.

\section{Stability Analysis of Fixed Points}
In this section, we conduct a stability analysis of individual plane wave solutions using regular phase plane techniques. This was already done for the alternative formulation of the traveling wave ODEs given in \cite{Saarloos}. We provide a brief derivation for our system \eqref{3.1.4a}, \eqref{3.1.4b} and \eqref{3.1.4c} for completeness and future use. However, a much more complex question is the issue of categorizing and elucidating the possible existence of, and transitions among, multiple plane wave states which may co--exist for the same parameter values in \eqref{3.1.1} (corresponding to the same operating conditions of the underlying system). Such behavior is well-documented in systems such as the Continuous Stirred Tank Reactor System \cite{Satsuma,Akhmediev:2}. For a system such as \eqref{3.1.1} and the associated ODEs \eqref{3.1.4a}, \eqref{3.1.4b} and \eqref{3.1.4c}, the large number of parameters makes a comprehensive parametric study of co--existing states bewilderingly complex, if not actually impracticable. 

We shall consider the cubic-quintic CGLE in the form \cite{Saarloos}
\begin{equation}\label{3.1.1}
\pa_tA=\epsilon A+(b_1+ic_1)\pa_x^2A-(b_3-ic_3) |A|^2A-(b_5-ic_5) |A|^4 A 
\end{equation}
noting that any three of the coefficients (no two of which are in the same term) may be set to unity by appropriate scalings of time, space and $A$. For normalized coefficients and when $c_1,c_3,c_5$ are small, authors \cite{Malo:2} have studied \eqref{3.1.1} in the context of a perturbed nonlinear Schr\"{o}dinger equation in which they found patterns that are stable with respect to small disturbances that triggered traveling wave state.

For the most part, we shall employ the polar form used in earlier treatments \cite{Malo:2,Brusch:1,Saarloos} of the traveling wave solutions of \eqref{3.1.1}. This takes the form of the ansatz
\begin{align}\label{3.1.2}
A(x,t) &= e^{-i\omega t} \hat A (x-vt)\notag\\
&= e^{-i\omega t} a(z)e^{i\phi (z)}
\end{align}
where
\begin{equation}\label{3.1.3}
z\equiv x-vt
\end{equation}
is the traveling wave variable and $\omega$ and $v$ are the frequency and translation speed (and are eigenvalues). Substitution of \eqref{3.1.2}/\eqref{3.1.3} in \eqref{3.1.1} leads, after some simplification, to the three mode dynamical system
\begin{subequations}\label{3.1.4}
\begin{align}
a_z &= b\label{3.1.4a}\\
b_z &= a\psi^2 -\gamma_1\Big[\gamma_2a+v\Big(b_1b+c_1\psi a\Big)-\gamma_3a^3-\gamma_4a^5\Big]\label{3.1.4b}\\
\psi_z &= -\frac{2\psi b}{a}+\gamma_1\Big[\gamma_5+v\Big(\frac{c_1b}{a}-b_1\psi\Big)-\gamma_6a^2-\gamma_7a^4\Big] \label{3.1.4c}
\end{align}
\end{subequations}
where $\psi\equiv\phi_z$. Note that we have put the equations into a form closer to that in \cite{Brusch:1}, rather than that in \cite{Saarloos}, so that \eqref{3.1.4} is a generalization of the traveling wave ODEs in \cite{Brusch:1} to include the quintic terms and the constant terms $\gamma_1-\gamma_7$ are given as functions of the system parameters in the following manner:
\begin{align}
\gamma_1&=\frac{1}{b_1^2+c_1^2}\nonumber\\
\gamma_2&=b_1\epsilon+c_1\omega\nonumber\\
\gamma_3&=b_1b_3-c_1c_3\nonumber\\
\gamma_4&=b_1b_5-c_1c_5\nonumber\\
\gamma_5&=-b_1\omega+c_1\epsilon  \nonumber\\
\gamma_6&=b_1c_3+c_1b_3\nonumber\\
\gamma_7&=b_1c_5+c_1b_5\nonumber\;.
\end{align}
From \eqref{3.1.2}, a fixed point $(a_0,0,\psi_0)$ of \eqref{3.1.4} corresponds to a plane wave solution
\begin{equation}\label{3.1.9}
A(x,t)=a_0e^{i(\psi_0z-\omega t)+i\theta}
\end{equation}
with $\theta$ an arbitrary constant. 

The fixed points of \eqref{3.1.4a}, \eqref{3.1.4b} and \eqref{3.1.4c} may be obtained by setting $b=0$ (from \eqref{3.1.4a}) in the right hand sides of the last two equations, solving the last one for $\psi$, and substituting this in the second yielding the quartic equation
\begin{equation}\label{3.1.10}
\alpha_4x^4+\alpha_3x^3+\alpha_2x^2+\alpha_1x+\alpha_0=0
\end{equation}
with
\begin{subequations}\label{3.1.11}
\begin{align}
x &= a^2,\label{3.1.11a}\\
\alpha_4&=\frac{\gamma_7^2}{b_1^2v^2}\label{3.1.11b}\\
\alpha_3&=\frac{2\gamma_6\gamma_7}{b_1^2v^2}\label{3.1.11c}\\
\alpha_2&=\frac{\gamma_6^2-2\gamma_5\gamma_7}{b_1^2v^2}+\frac{\gamma_1(b_1\gamma_4+c_1\gamma_7)}{b_1}\label{3.1.11d}\\
\alpha_1&=\gamma_1\big(\gamma_3+\frac{c_1\gamma_6}{b_1}\big)-\frac{2\gamma_5\gamma_6}{b_1^2v^2} \label{3.1.11e}\\
\alpha_0&=\frac{\gamma_5^2}{b_1^2v^2}-\frac{\gamma_1}{b_1}\big(b_1\gamma_2+c_1\gamma_5\big)
\;.\label{3.1.11f}
\end{align}
\end{subequations}
Thus, with $a_0=\sqrt{x}$ for $x$ any of the four roots of \eqref{3.1.10}, we have a plane wave solution of the form \eqref{3.1.9}.
For each of the four roots $x_i,i=1, \dots ,4$ of \eqref{3.1.10} corresponding to a fixed point of \eqref{3.1.4a}, \eqref{3.1.4b} and \eqref{3.1.4c} or a plane wave $\sqrt{x_i} \; e^{i(\psi_iz-wt)+i\theta_i}$, the stability may be determined using regular phase--plane analysis. The characteristic polynomial of the Jacobian matrix of a fixed point $x_i=a_i^2$ of \eqref{3.1.4a}, \eqref{3.1.4b} and \eqref{3.1.4c} may be expressed as
\begin{equation} \label{3.1.12}
\lambda^3+\delta_1\lambda^2+\delta_2\lambda +\delta_3=0
\end{equation}
where
\begin{subequations}
\begin{align}
\delta_1&=2b_1v\gamma_1 \label{3.1.13a}\\
\delta_2&=3\psi^2+\gamma_1[\gamma_2-a^2(3\gamma_3+5a^2\gamma_4)-v(3c_1\psi -v)]\label{3.1.13b}\\
\delta_3&=-2a^2\gamma_1(\gamma_6+2a^2\gamma_7)(-2\psi+c_1\gamma_1v)\notag\\
&+b_1\gamma_1v[-\psi^2+\gamma_1(\gamma_2-3a^2\gamma_3-5a^4\gamma_4+c_1\psi v)]\label{3.1.13c}
\end{align}
\end{subequations}
where the fixed point values $(a_i,\psi_i)=(\sqrt{x_i}, \psi_i)$ are to be substituted in terms of the system parameters. Note that $\psi_i$ is obtained by setting $a=a_i=\sqrt{x_i},$ and $ b=0$ in the right side of \eqref{3.1.4c}.

For $(a_0,0,\psi_0)$ to be a stable fixed point within the linearized analysis, all the eigenvalues must have negative real parts. Using the Routh--Hurwitz criterion, the necessary and sufficient conditions for \eqref{3.1.12} to have $Re(\lambda_{1,2,3})<0$ are:
\begin{equation}\label{3.1.14}
\delta_1>0,\quad\delta_3>0,\quad\delta_1\delta_2-\delta_3>0.
\end{equation}
Equation \eqref{3.1.14} is thus the condition for stability of the plane wave corresponding to $x_i$.

On the contrary, one may have the onset of instability of the plane wave solution occurring in one of two ways. In the first, one root of \eqref{3.1.12} (or one eigenvalue of the Jacobian) becomes non--hyperbolic by going through zero for 
\begin{equation} \label{3.1.15}
\delta_3=0.
\end{equation}
Equation \eqref{3.1.15} is thus the condition for the onset of ``static'' instability of the plane wave. Whether this bifurcation is a pitchfork or transcritical one, and its subcritical or supercritical nature, may be readily determined by deriving an appropriate canonical system in the vicinity of \eqref{3.1.15} using any of a variety of normal form or perturbation methods \cite{Drazin:2,Murray,Balmforth}.

One may also have the onset of dynamic instability (``flutter'' in the language of Applied Mechanics) when a pair of eigenvalues of the Jacobian become purely imaginary. The consequent Hopf bifurcation at
\begin{equation}\label{3.1.16}
\delta_1\delta_2-\delta_3=0
\end{equation}
leads to the onset of periodic solutions of \eqref{3.1.4a}, \eqref{3.1.4b} and \eqref{3.1.4c} (dynamic instability or ``flutter''). These periodic solutions for $a(z)$ and $\psi (z)$, which may be stable or unstable depending on the super-- or subcritical nature of the bifurcation, correspond via \eqref{3.1.2} to solutions
\begin{equation}\label{3.1.17}
A(x,t)=a(z)e^{i\left(\int \psi dz-\omega t\right)}
\end{equation}
of the CGLE \eqref{3.1.1} which are, in general, quasiperiodic wavetrain solutions. This is because the period of $\psi$ and $\omega$ are typically incommensurate. Eq. \eqref{3.1.17} is periodic if $\omega =0$. 

\section{Stability Analysis of Periodic Orbits}
In this section we will use the method of multiple scales to construct analytical approximations for the periodic orbits arising through Hopf bifurcation of the fixed point of the CGLE equation. For the systems of differential equations given by \eqref{3.1.4a}, \eqref{3.1.4b} and \eqref{3.1.4c}, the physically relevant point is given by $(a_0,0,\psi_0)$  where $\psi_i$ is obtained by setting $a=a_i=\sqrt{x_i},$ in
\begin{equation}\label{3.1.18}
\psi_i=\frac{\gamma_5-a_i^2(\gamma_6+a_i^2\gamma_7)}{b_1v}
\end{equation}
and $x_i$ is one of the roots of the fixed point equation \eqref{3.1.10}.
We will choose the parameter $\epsilon$ which represents the linear gain or loss as the control parameter. The limit cycle is determined by expanding about the fixed point using progressively slower time scales. The expansion takes the form
\begin{align}
a&=a_0+\sum_{n=1}^{3}\delta^n a_n(Z_0,Z_1,Z_2)+\cdots \label{3.2.1},\\
b&=B_0+\sum_{n=1}^{3}\delta^n B_n(Z_0,Z_1,Z_2)+\cdots\label{3.2.2},\\
\psi&=\psi_0+\sum_{n=1}^{3}\delta^n \psi_n(Z_0,Z_1,Z_2)+\cdots,\label{3.2.3}
\end{align}
where $Z_n=\delta^n$ t and $\delta$ is a small positive non-dimensional parameter that is introduced as a bookkeeping device and will be set to unity in the final analysis. Using the chain rule, the time derivative becomes
\begin{equation}\label{3.2.4}
\frac{\mathrm{d}}{\mathrm{d}Z}=D_0+\delta D_1+\delta^2 D_2+\cdots,
\end{equation}
where $D_n=\partial/\partial Z_n$. The delay parameter $\epsilon$ is ordered as 
\begin{equation}\label{3.2.5}
\epsilon=\epsilon_0+\delta^2\epsilon_2,
\end{equation}
where $\epsilon_0$ is the critical value such that \eqref{3.1.14} is not satisfied, (i.e. $\epsilon_0$ is a solution of
\eqref{3.1.16}). This is standard for this method, as it allows the influence from the nonlinear terms and the control parameter to occur at the same order.

Using \eqref{3.2.1}--\eqref{3.2.5} in \eqref{3.1.4a}--\eqref{3.1.4c} and equating like powers of $\delta$ yields equations at $\mathrm{O}(\delta^i)$, $i=1,2,3$ of the form:
\begin{align}
L_1(a_i,B_i,\psi_i)&=S_{i,1}\label{3.2.6},\\
L_2(a_i,B_i,\psi_i)&=S_{i,2}\label{3.2.7},\\
L_3(a_i,B_i,\psi_i)&=S_{i,3}\label{3.2.8},
\end{align}
where, the $L_i$, $i=1,2,3$ are the differential operators
\begin{align}
L_1(a_i,B_i,\psi_i)&=D_0a_i-B_i\equiv S_{i,1}\label{3.2.9},\\
L_2(a_i,B_i,\psi_i)&=D_0B_i-\psi_0^2a_i-2a_0\psi_0\psi_i\notag\\
&+\gamma_1\{\gamma_{20}a_i+v[b_1B_i+c_1(\psi_0a_i+a_0\psi_i)]\notag\\
&-3\gamma_{30}a_0^2a_i-5\gamma_{40}a_0^4a_i\}\equiv S_{i,2}\label{3.2.10},\\
L_3(a_i,B_i,\psi_i)&=a_0(D_0\psi_i)+2(\psi_0B_i+B_0\psi_i)\notag\\
&-\gamma_1\{\gamma_{50}a_i+v[c_1B_i-b_1(\psi_0a_i+a_0\psi_i)]\notag\\
&-3\gamma_{60}a_0^2a_i-5\gamma{70}a_0^4a_i\}\equiv S_{i,3}\label{3.2.11},
\end{align}
where $\gamma_p=\gamma_{p0}+\delta^2\gamma_{p2}$ with $p=2,3,\cdots,7$, the source terms $S_{i,j}$ for $i,j=1,2,3$ at $\mathrm{O}(\delta)$, $\mathrm{O}(\delta^2)$, and $\mathrm{O}(\delta^3)$ are given by the following:
$$\mathrm{O}(\delta):$$
\begin{align}
S_{1,1}&=0 \label{A.1}\\
S_{1,2}&=0 \label{A.2}\\
S_{1,3}&=0 \label{A.3}.
\end{align}
$$\mathrm{O}(\delta^2):$$
\begin{align}
S_{2,1}&=-D_1a_1 \label{A.4}\\
S_{2,2}&=-D_1B_1+a_0\psi_1^2+2\psi_0a_1\psi_1-\gamma_1(\gamma_{22}a_0+vc_1\psi_1a_1-3\gamma_{30}a_0a_1^2 \notag\\
&-\gamma_{32}a_0^3-10\gamma_{40}a_0^3a_1^2-\gamma_{42}a_0^5) \label{A.5}\\
S_{2,3}&=-a_0D_1\psi_1-a_1D_0\psi_1-2\psi_1B_1+\gamma_1[(\gamma_{52}a_0-vb_1\psi_1a_1)\notag\\
&-(3\gamma_{60}a_0a_1^2+\gamma_{62}a_0^3)-(10\gamma_{70}a_0^3a_1^2+\gamma_{72}a_0^5)] \label{A.6}.
\end{align}
$$\mathrm{O}(\delta^3):$$
\begin{align}
S_{3,1}&=-D_1a_2-D_2a_1 \label{A.7}\\
S_{3,2}&=-D_1B_2-D_2B_1+2a_2\psi_0\psi_1+a_1(2\psi_0\psi_2+\psi_1^2)+2a_0\psi_1\psi_2\notag\\
&-\gamma_1\{\gamma_{22}a_1+vc_1(\psi_1a_2+\psi_2a_1)-[\gamma_{30}(a_1^3+6a_0a_1a_2)+3\gamma_{32}a_0^2a_1] \notag\\
&-[\gamma_{40}(10a_0^2a_1^3+20a_0^3a_1a_2)+5\gamma_{42}a_0^4a_1]\} \label{A.8}\\
S_{3,3}&=-D_1\psi_2-D_2\psi_1-a_1(D_1\psi_1+D_0\psi_2)-a_2D_0\psi_1-2(\psi_1B_2+\psi_2B_1)\notag\\
&+\gamma_1\{\gamma_{52}a_1-vb_1(\psi_1a_2+\psi_2a_1)-[\gamma_{60}(a_1^3+6a_0a_1a_2)\notag\\
&+3\gamma_{62}a_0^2a_1]-[\gamma_{70}(10a_0^2a_1^3+20a_0^3a_1a_2)+5\gamma_{72}a_0^4a_1]\} \label{A.9}.
\end{align}

Also, \eqref{3.2.6} may be solved for $B_i$ in terms of $a_i$. and $\psi_i$. Using this in \eqref{3.2.7} yields $\psi_i$
\begin{equation}\label{3.2.12}
\psi_i=\frac{\theta_i}{\phi_1},
\end{equation}
where
\begin{align}\label{3.2.13}
\theta_i&=-D_0S_{i,1}+D_0^2a_i-\psi_0^2+\gamma_1\{\gamma_{20}a_i-3\gamma_{30}a_0^2a_i-5\gamma_{40}a_0^4a_i\notag\\
&+v[b_1(-S_{i,1}+D_0a_i)+c_1]\}-S_{i,2}
\end{align}
and 
\begin{equation}\label{3.2.14}
\phi_1=2a_0\psi_0-v\gamma_1c_1a_0.
\end{equation}

Using \eqref{3.2.12} and the equation for $B_i$  in \eqref{3.2.9} yields the composite equation:
\begin{equation}\label{3.2.15}
L_ca_i\equiv\Gamma_i,
\end{equation}
where
\begin{align}
\Gamma_i&\equiv S_{i,3}-\frac{a_0}{\phi_1}\Big(D_0\zeta_i\Big)-\frac{2B_0}{\phi_1}\zeta_i-\gamma_1vb_1a_0\frac{\zeta_i}{\phi_1}+(2\psi_0-\gamma_1vc_1)S_{i,1}, \label{3.2.16}\\
\zeta_i&=-D_0S_{i,1}-\gamma_1vb_1S_{i,1}-S_{i,2}. \label{3.2.17}
\end{align}

We shall now use \eqref{3.2.16} and \eqref{3.2.17} to systematically identify and suppress secular terms in the solutions of \eqref{3.2.9},\eqref{3.2.10},\eqref{3.2.11}.
Let us now turn to finding the solutions of \eqref{3.2.9},\eqref{3.2.10},\eqref{3.2.11}. In what follows , we shall detail the solution of the above system of equations for the case $\epsilon_0=\epsilon_{01}$. In order to achieve that we must find first the fixed points. The characteristic polynomial of the Jacobian matrix of a fixed point of \eqref{3.1.4a},\eqref{3.1.4b},\eqref{3.1.4c} may be expresses as 
\begin{equation} \label{3.2.18}
\lambda^3+\delta_1\lambda^2+\delta_2\lambda +\delta_3=0,
\end{equation}
as in \eqref{3.1.12}, and the fixed point values $(a_i,\psi_i)$ are to be substituted in terms of the system parameters.

The condition $\delta_1\delta_2-\delta_3=0$ yields an involved equation in $\epsilon$ which actually can be solved easily numerically for $\epsilon_0$ by the root method .

For $\mathrm{O}(\delta)$  the Eqns. \eqref{A.1}--\eqref{A.3} give $S_{i,1}=S_{i,2}=S_{i,3}=0$, and hence we may pick a solution for the first order as
\begin{equation}\label{3.2.20}
a_1=\alpha(Z_1,Z_2)e^{\lambda_1 Z}+\beta(Z_1,Z_2)e^{\lambda_2 Z}+\gamma(Z_1,Z_2)e^{\lambda_3 Z},
\end{equation}
where $\beta=\bar{\alpha}$  is the complex conjugate of $\alpha$ and $\lambda_2=\lambda_1$. As evident for the Routh--Hurwitz condition, the $\alpha$ and $\beta$ modes correspond to the center manifold where $\lambda_{1,2}$ are purely imaginary and where the Hopf bifurcation occurs, while $\gamma$ corresponds to the attractive direction or the stable manifold. Since we wish to construct and analyze the stability of the periodic orbits which lie in the center manifold, we should take $\gamma=0$ so \eqref{3.2.20} becomes
\begin{equation}\label{3.2.21}
a_1=\alpha(Z_1,Z_2)e^{i\omega Z}+\beta(Z_1,Z_2)e^{-i\omega Z}.
\end{equation}
Using \eqref{A.1}--\eqref{A.3} for $i=1$ then the first order fields $(a_1,B_1,\psi_1)$ are 
\begin{equation}\label{3.2.22}
B_1=D_0a_1=i\omega\alpha e^{\imath\omega Z}-i\omega\beta e^{-i\omega Z},
\end{equation}
and also \eqref{3.2.12} becomes
\begin{align}\label{3.2.23}
\psi_1&=\frac1{\phi_1}\Big[-\omega^2-\psi_0^2+\gamma_1\Big(\gamma_{20}+vc_1\psi_0-3\gamma_{30}a_0^2-5\gamma_{40}a_0^4\Big)\Big]\notag\\
&\times \Big(\alpha e^{i\omega Z}+\beta e^{-i\omega Z}\Big)+\frac{\gamma_1vb_1}{\phi_1}\Big(i\omega \alpha e^{i\omega Z}-i\omega \beta e^{-i\omega Z}\Big).
\end{align} 

Now that the first order solutions \eqref{3.2.21}--\eqref{3.2.23} are known, the second order sources $S_{21}$, $S_{22}$, $S_{23}$ may be evaluated via \eqref{A.4}--\eqref{A.6}. Using these sources in \eqref{3.2.16} we obtain $\Gamma_2$ which may be written as 
\begin{equation}\label{3.2.24}
\Gamma_2=\Gamma_2^{(0)}+\Gamma_2^{(1)}e^{i\omega Z}+\Gamma_2^{(2)}e^{2i\omega Z}+c.c.
\end{equation}
Setting the coefficients of the secular $e^{i\omega Z}$ terms (which are the solutions of the homogeneous equation for $i=1$) to zero, i.e. $\Gamma_2^{(1)}=0$ yields
\begin{align}\label{3.2.25}
D_1\alpha&=\frac{\pa\alpha}{\pa{Z_1}}=0,\notag \\
D_1\beta&=\frac{\pa\beta}{\pa{Z_1}}=0.
\end{align}
Using \eqref{3.2.25}, the second order sources, and assuming a second-order particular solution for $a_2$ of the form:
\begin{equation}\label{3.2.26}
a_2=a_2^{(0)}+a_2^{(2)}e^{2i\omega Z},
\end{equation}
having the standard form of a DC or time--independent term plus second--harmonic terms, the composite equations \eqref{3.2.15}--\eqref{3.2.17} for $i=2$, yield 
\begin{equation}\label{3.2.27}
L_ca_2=\Gamma_2^{(0)}+\Gamma_2^{(2)}e^{2i\omega Z},
\end{equation}
which will be solved for the particular solution $a_2^{(0)}$, and $a_2^{(2)}$ by equating both sides of the expression \eqref{3.2.27}. In terms of the operator $L_c$ which is obtained from \eqref{3.2.16}, the particular solution takes the form:
\begin{align}
a_2^{(0)}&=-\Gamma_2^{(0)}\big[a_0(vc_1\gamma_1-2\psi_0)\big]\Big\{2B_0\big[-\psi_0^2+\gamma_1(\gamma_{20}-3a_0^2\gamma_{30}-5a_0^4\gamma_{40}+vc_1\psi_0)\big]\notag\\
&+a_0\gamma_1\big\{(\gamma_{50}-3a_0^2\gamma_{60}-5a_0^4\gamma_{70})(vc_1\gamma_1-2\psi_0)\notag\\
&+vb_1[\gamma_1(\gamma_{20}-3a_0^2\gamma_{30}-5a_0^4\gamma_{40})+\psi_0^2]\big\}\Big\}^{-1},\\
a_2^{(2)}&=-\Gamma_2^{(2)}\big[a_0(vc_1\gamma_1-2\psi_0)\big]\Big\{6a_0^2B_0\gamma_1\gamma_{30}+10a_0^4B_0\gamma_1\gamma_{40}\notag\\
&+3a_0^3\gamma_1\big[\gamma_{30}(2i\omega+vb_1\gamma_1)+\gamma_{60}(vc_1\gamma_1-2\psi_0)\big]\notag\\
&+5a_0^5\gamma_1\big[\gamma_{40}(2i\omega+vb_1\gamma_1)+\gamma_{70}(vc_1\gamma_1-2\psi_0)\big]\notag\\
&+2B_0\big[4\omega_2-2iv\omega b_1\gamma_1+\psi_0^2-\gamma_1(\gamma_{20}+vc_1\psi_0)\big]\notag\\
&+a_0\big\{8i\omega^3-2iv^2\omega b_1^2\gamma_1^2-6i\omega\psi_0^2+2\gamma_1\psi_0(3iv\omega c_1+\gamma_{50})\notag\\
&+\gamma_1[-2i\omega(v^2c_1^2\gamma_1+\gamma_{20})-vc_1\gamma_1\gamma_{50}]-vb_1\gamma_1(-8\omega_2+\gamma_1\gamma_{20}+\psi_0^2)\big\}\Big\}^{-1}.\label{3.2.28}
\end{align}
Using \eqref{3.2.12}, the second order sources and the equation for $B_i$  in \eqref{3.2.9} with $i=2$ then we can find the second order fields $B_2$ and $\psi_2$. Substituting them into the \eqref{A.7}--\eqref{A.9} we find the third order sources and we may evaluate the coefficients of the secular term $e^{i\omega t}$ in the composite source $\Gamma_3$ of \eqref{3.2.16}. Suppressing again the secular terms to obtain uniform expansions yields the final equation for the evolution of the coefficients in the linear solutions \eqref{3.2.21}-\eqref{3.2.23} on the slow second--order time scales
\begin{equation}\label{3.2.29}
\frac{\pa\alpha}{\pa{Z_2}}=S_1\alpha^2\beta+S_2\alpha.
\end{equation}
Writing $\alpha=\frac12Ae^{i\theta}$ and separating \eqref{3.2.29} into real and imaginary parts, yields
\begin{equation}\label{3.2.30}
\frac{\pa A}{\pa{Z_2}}=\frac{S_{1r}A^3}{4}+S_{2r}A,
\end{equation}
where $S_{1r}$ and $S_{2r}$ represent the real parts of $S_1$ and $S_2$ respectively.
In the usual way, the fixed points of \eqref{3.2.30}, $(A_1,A_{2,3})$ where
\begin{align}
A_1&=0,\notag\\
A_{2,3}&=\pm 2\sqrt{-\frac{S_{2r}}{S_{1r}}} \label{3.2.31}
\end{align}
give the amplitude of the solution $\alpha=\frac12Ae^{i\theta}$, with $A_{2,3}$ corresponding to the bifurcation periodic orbits. Clearly $A_{2,3}$ are real fixed points whenever 
\begin{equation}\label{3.2.32}
\frac{S_{2r}}{S_{1r}}<0,
\end{equation}
and the Jacobian of the right hand side of \eqref{3.2.32} evaluated at $A_{2,3}$ is $J|_{A_{2,3}}=-S_{2r}$. Clearly, a necessary condition for stability is to have $S_{2r}>0$, and for instability $S_{2r}<0$. Thus, the system undergoes:
\begin{itemize}
\item[a.] supercritical Hopf bifurcations when 
\begin{equation}
S_{2r}>0, \qquad S_{1r}<0,
\end{equation}
\item[b.]	subcritical Hopf bifurcations when
\begin{equation}
S_{2r}<0, \qquad S_{1r}>0.
\end{equation}
\end{itemize}

\section{Discussion of Results}

In this section, we consider the numerical results which follow from the analysis in the previous section. The fixed point equation \eqref{3.1.10} can be solved analytically for each fixed point $x_i$ using the program Mathematica, for $i=1,\cdots,4$. The characteristic polynomial of the Jacobian matrix of a fixed point of \eqref{3.1.4a},\eqref{3.1.4b},\eqref{3.1.4c} may be expresses as is \eqref{3.1.18}. Since all coefficients $\alpha_i$, for $i=1\cdots3$ depend on the nine system parameters, we fix $b_1=0.08$, $b_3=-0.65$, $b_5=0.1$, $c_1=0.5$, $c_3=1$, $c_5=-0.07$, $\omega=0$, and $v=0.01$. The possibility of bounded chaotic solitons depends on the system being fairly strongly dissipative near the fixed points $(a_0,0,\psi_0)$ in a significant part of the phase space, with the strong dissipativity ruling out the appreciable  volume expansion associated with an attractor at infinity, as well as volume--conserving quasiperiodic behavior. The trace of the Jacobian matrix for this sets of values at the fixed point $(a_0,0,\psi_0)$, which gives the local logarithmic rate of change of $(a,b,\psi)$ phase--space volume $V$ is
$\frac1V\frac{\mathrm{d}v}{\mathrm{d}t}=J(a_0,0,\psi_0)=-0.0062$, so we may anticipate that the orbits may go to an attractor at infinity, since the dissipation is weak.

The four fixed points can be analytically found as a function of only one parameter, in our case we chose $\epsilon$ as being the free parameter. By choosing ``the right fixed point'', the Hopf curve $\alpha_1\alpha_2-\alpha_3=0$ may be solved numerically for $\epsilon$, which gives $\epsilon_0=-0.0000807$. The idea is to find the ``right'' $\epsilon$ which will give rise to the condition for Hopf bifurcation, (i.e. $\alpha_1>0$, $\alpha_2>0$, $\alpha_3>0$ and $\alpha_1\alpha_2-\alpha_3<0$).

We obtain $\alpha_1=0.006$, $\alpha_2=0.001$, $\alpha_3=0.0001$ and $\alpha_1\alpha_2-\alpha_3=-1.01\ 10^{-6}$ for an $\epsilon=-0.008$, i.e.  $\epsilon<\epsilon_0$.

Now we will analyze the multiple scales method to construct the analytical approximations for the periodic orbits arising through the Hopf bifurcations of the fixed point. The delay parameter $\epsilon$ (or the bifurcation parameter) is ordered as $\epsilon=\epsilon_0+\delta^2\epsilon_2$, where $\epsilon_0=-0.0000807$, and $\epsilon_2=-0.1$. This method allows the influence from the nonlinear terms and the control parameter to occur at the same order. For the system parameters chosen above, at the fixed points, we get $(a_0,0,\psi_0)=(0.0121663,0,-0.00514)$.

From \eqref{3.2.15} and by the method presented in Section 2, the final equation for the evolution coefficients in the linear solutions, on the slow second--order time scale is
\begin{equation}\label{3.3.1}
\frac{\pa\alpha(Z_1,Z_2)}{\pa{Z_2}}=S_1\alpha^2(Z_1,Z_2)\beta(Z_1,Z_2)+S_2\alpha(Z_1,Z_2),
\end{equation}
where $S_1=-3235.55+295.279 i$ and $S_2=297.074-32.26i$. Since $S_{2r}=Re(S_2)>0$, and $S_{1r}=Re(S_1)<0$, then this situation will correspond to a supercritical Hopf bifurcation. 
\begin{figure}
\begin{center} 
\includegraphics[width=350pt]{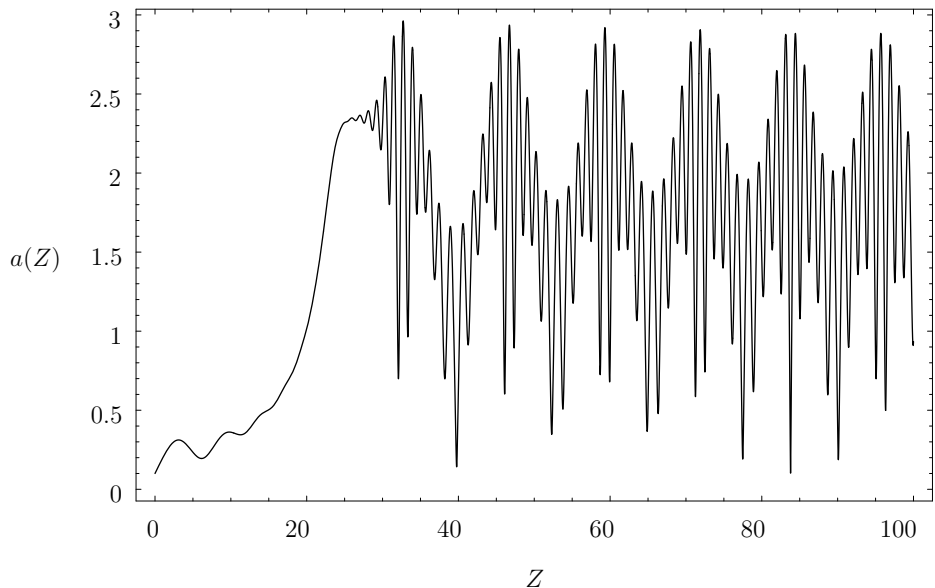}
\end{center}
\caption{Stable periodic oscillations on the limit cycle} \label{Figure 33}
\end{figure}
Figure \ref{Figure 33} shows the time behavior for $a(Z)$ for $\epsilon=-0.00008$ (the supercritical regime). Note that, as anticipated from before, there is a stable limit cycle attractor at $\epsilon$, the solution remains positive and bounded while it stays periodic. 

Clearly, similar stable periodic solutions may be obtained for many other parameter sets. For each case, the overall solution $A(x,t)$ of the CGLE is, via \eqref{3.1.17}, a quasiperiodic solution.

One may also use the above approach to both explain, and extend, the numerical treatment by Brusch et al \cite{Brusch:1,Brusch:2} of the periodic traveling waves of the CGLE using the bifurcation software AUTO. However, the solutions in Brusch et al do not appear to be clearly correlated to the dissipative solitons of the CGLE in Akhmediev at al \cite{Akhmediev:2}. Hence, we shall move on next to briefly consider possible generalizations of the above treatment.

\section{Remarks on Generalized Hopf Bifurcations and Emergence of Global Structure}
One may pursue the line of inquiry based on the traveling waves or spatial ODEs even further to track the emergence of global structure. We have done preliminary work along these lines. However, although there is a well--established roadmap and it has been implemented in detail for the well known Continuous Stirred Tank Reactor System \cite{Golubitski,Planeaux}, we are not convinced of its relevance to the actual numerical simulations of dissipative solitons \cite{Akhmediev:2,Akhmediev:1}. Hence, we present it here as a possible future direction to pursue.

For completeness, let us first consider more degenerate cases where more than one root of the Jacobian is non--hyperbolic. In such cases the non--hyperbolic eigenvalues of the Jacobian matrix, may consist of either:
\begin{itemize}
\item[a.]a double zero: \qquad $\lambda_{1,2}=0$  \qquad $\lambda_3\in\Re$
\item[b.]one zero and a complex conjugate pair: \qquad $\lambda_1=0$  \qquad $\lambda_2=\bar{\lambda}_3$
\item[c.]a triple zero: \qquad $\lambda_{1,2,3}=0$
\end{itemize}
For the above situations, we have the following sub--cases of the so--called ``degenerate Hopf'' $(H1)$ bifurcation. Each sub--case is given a name:
\begin{equation}
\mathrm{F_1:} \qquad \lambda_{1,2}=0,0  \label{4.0.6}
\end{equation}
\begin{equation}
\mathrm{F_2:} \qquad \lambda_{1,2,3}=\pm i\omega_0,0  \label{4.0.7}
\end{equation}
\begin{equation}
\mathrm{G_1:} \qquad \lambda_{1,2,3}=0,0,0  \label{4.0.8}
\end{equation}
In these cases, \cite{Planeaux,Arnold,Giuckenhemer} and \cite{Langiford,Pismen:1,Pismen:2}, these $(H1)$ bifurcations may lead to global structure including homoclinic orbits, invariant tori, and period doubling to chaos at the $(H1)$ points. One may also work perturbatively \cite{Arnold} near these $(H1)$ points as done by Keener for the well--known Continuous Stirred Tank Reactor problem. 

Two other degenerate/generalized Hopf bifurcation scenarios are possible. As seen in Chapter 3 \eqref{3.2.30}, the normal form for the Hopf bifurcation may be written as
\begin{equation}
\dot{r}=r\big[\alpha(\mu)+c_1(\mu)r^2+c_2(\mu)r^4+...\big] \label{4.0.1}
\end{equation}
\begin{equation}
\dot{\theta}=\omega_0 +O(\mu,r^2)  \label{4.0.2}
\end{equation}
where we have made the identification $A\rightarrow r$, $S_{1r}/4\rightarrow c_1$,  $S_{2r}\rightarrow \alpha$, and higher order nonlinear terms are included.

The first kind of possible degeneracy (the $(H2)$ kind) occurs if 
\begin{align}\label{4.0.3}
&\alpha=\alpha'=...=\alpha^{(k)}=0 \notag\\
&\alpha^{(k+1)}\neq0.
\end{align}
This is the so--called $\mathrm{k^{th}}$ order $(H2)$ degeneracy and it gives rise to multiple Hopf points and multiple periodic orbits. The resulting structure is thus similar to that resulting from a regular Hopf bifurcation, and much less complex than the structure produced by $(H1)$ bifurcation.

A second possible degeneracy in the normal form \eqref{4.0.1} corresponds to
\begin{align}\label{4.0.4}
&c_1=c_2=...=c_m=0 \notag\\
&c_{m+1}\ne0.
\end{align}
This so--called $\mathrm{m^{th}}$ order $(H3)$ degeneracy results in isolated branches of periodic solutions unconnected to the main branch.

When the $\mathrm{k^{th}}$ order $(H2)$ degeneracy and the $\mathrm{m^{th}}$ order $(H3)$ degeneracy occur simultaneously, the normal form  \eqref{4.0.1} may be rescaled to the form:
\begin{equation}
\dot{r}=r\big[r^{2m+2}+...\pm \mu^{k+1}\big] \label{4.0.5}
\end{equation}
This is the so--called $H_{mk}$ degeneracy.

In the case of the $(H2)$ degeneracy, the complex conjugate eigenvalues $\pm i\omega$ at the Hopf point cross the imaginary axis tangentially leading, after additional analysis, to multiple periodic orbits.

For $(H3)$ degeneracy, one may obtain isolated branches (isolas) of periodic orbits unconnected to the main branch.

However, of greatest interest are the $(H1)$ bifurcations where the Jacobian has more than one non--hyperbolic eigenvalue and global structure emerges. These will be pursued in future work, together with the $(H2)$ and $(H3)$ cases.

\bibliographystyle{plain} %% {paper}
\bibliography {diss}    % bibliography references

\end{document}